\documentclass[useAMS,usenatbib]{mn2e}

\usepackage{amsmath,psfig,epsfig,graphics}

\def\apj{ApJ}

\def\apjl{ApJ}
\def\mnras{MNRAS}

\def\aj{AJ}

\def\physrep{Phys. Rep.}
\def\nat{Nat}

\def\prd{Phys.~Rev.~D}
\def\pasj{PASJ}

\def\lesssim{\mathrel{\hbox{\rlap{\hbox{\lower4pt\hbox{$\sim$}}}\hbox{$<$}}}}
\def\gesssim{\mathrel{\hbox{\rlap{\hbox{\lower4pt\hbox{$\sim$}}}\hbox{$>$}}}}

\def\lesssim{\mathrel{\hbox{\rlap{\hbox{\lower4pt\hbox{$\sim$}}}\hbox{$<$}}}}
\def\gesssim{\mathrel{\hbox{\rlap{\hbox{\lower4pt\hbox{$\sim$}}}\hbox{$>$}}}}

\newcommand\ion[2]{#1$\;${\small\rmfamily\@Roman{#2}}\relax}%

\begin{document}

\author[A. Morandi and R. Barkana]
{Andrea Morandi${}^1$\thanks{E-mail: andrea@wise.tau.ac.il} and Rennan Barkana${}^{1}$\\
$^{1}$ School of Physics and Astronomy, Tel Aviv University, Tel Aviv, 69978, Israel\\
}



\title[Studying reionization with global 21-cm]{Studying cosmic reionization with observations of the global 21-cm signal}
\maketitle

\begin{abstract}
We explore the ability of observations of the global brightness
temperature of the 21-cm signal to constrain the reionization history
and the properties of the ionizing sources. In order to describe the
reionization signal, we employ either a commonly-used toy model or a
more realistic structure formation model that parameterizes the
properties of the ionizing sources. If the structure formation model
captures the actual evolution of the reionization signal, then
detecting the signal is somewhat easier than it would be for the toy
model; using the toy model in this case also leads to systematic
errors in reconstructing the reionization history, though a
sufficiently sensitive experiment should be able to distinguish
between the two models. We show that under optimistic assumptions
regarding systematic noise and foreground removal, one-year
observations of the global 21-cm spectrum should be able to detect a
wide range of realistic models and measure the main features of the
reionization history while constraining the key properties of the
ionizing sources.
\end{abstract}

\begin{keywords}
galaxies:high-redshift -- cosmology:theory -- galaxies:formation
\end{keywords}

\section{Introduction}\label{intro}
One of the most important frontier fields of cosmology is the
evolution of the Universe from the dark ages following hydrogen
recombination through to the epoch of reionization. The 21-cm line
associated with the hyperfine transition of atomic hydrogen is the
most promising signal for detecting and mapping the spatial and
redshift distribution of hydrogen in the universe, and for studying
the sources responsible for heating and reionizing the intergalactic
medium (IGM) at redshifts $z\gesssim 6$. Indeed, an important feature of
the 21-cm signal is that the spectral dimension allows in principle 3D
tomography of hydrogen as a function of redshift, providing much
richer structure than the cosmic microwave background (CMB), which
yields just a single sky map; this may help to detect primordial
non-Gaussianity and test inflation \citep{loeb2004}. During
reionization, detecting the bubble structure would probe the main
sources of ionizing radiation, even if these are otherwise
unobservable because, for example, they are too faint to be detected
individually. It is also important to characterize when and how long
the reionization took place, because of the significant effect of
reionization on the subsequent formation of galaxies. In particular,
the ultra-violet (UV) radiation heats the gas, raising the Jeans mass
and causing a suppression of star-forming galaxies in low-mass halos
($\lesssim 10^9 M_{\odot}$).

A first estimate of the reionization redshift $z_r$ has been deduced
from the CMB polarization, where an additional peak on large angular
scales, corresponding to the horizon size at the reionization epoch,
is expected due to scattering, with an amplitude related to the total
optical depth. The recent analysis based on the 7-year WMAP data
\citep{komatsu2010} finds a reionization redshift
of $10.4\pm1.2$, in terms of an equivalent instantaneous reionization,
but is still consistent with a wide range of possible reionization
histories. Other constraints come from the Ly$\alpha$ galaxies at
redshift 5.7 and 6.5, whose characteristic luminosity function shows a
lack of time evolution that is consistent with a fully ionized IGM at
$z \sim 6$ \citep{malhotra2004}.  Analyses of the spectra of 
high-z QSOs
\citep{fan2004, goto2006, willott2007, willott2009, mortlock2011} 
and Gamma Ray Bursts (GRB) \citep{totani2006} also suggest that the
IGM is still very highly ionized at this redshift.

The expected picture for reionization is thus an inhomogeneous and
extended process, for which the nature and the evolution of the
ionizing sources are still observationally undetermined. Upcoming and
future observational probes should allow us to distinguish among
various reionization models and in particular constrain the possible
extended or instantaneous nature of this process
\citep{bruscoli2002}. Low-frequency observations with radio telescope arrays
such as LOFAR\footnote{http://www.lofar.org/},
MWA\footnote{http://www.MWAtelescope.org/},
PAPER\footnote{\citet{parsons2010}}, and
SKA\footnote{http://www.skatelescope.org/}, will over the next decade
constrain the spatial distribution of the ionization sources, while
dipole observations of the spatially-integrated 21-cm signal are
currently underway, although in their infancy, as exemplified by EDGES
\citep{bowman2008,BowmanNature}.

Recently \citet[][hereafter PL]{pritchard2010} made a first
theoretical attempt to predict detection limits for future
observations of the global 21-cm signal. In this paper, we explore the
potential for global 21-cm experiments to constrain reionization
starting with the same simple analytical toy model used by PL (based
on the {\em tanh} function), but then focus on a more realistic,
physically-based galaxy formation model that parameterizes the
properties of the ionizing sources. In the following section we
briefly review the general setup of global 21-cm measurements and the
toy model, and then introduce a simple galaxy formation model within
Cold Dark Matter (CDM) - dominated hierarchical structure
formation. In section~3 we make predictions for the
spatially-integrated 21-cm signal that experiments such as EDGES aim
to measure. We compare the expected signal from the two models, and
show some examples of the expected errors in the model parameters that
are reconstructed from observations. In section~3.4 we explore the
systematic effects of assuming an incorrect model when trying to
reconstruct the global 21-cm signal. We finish this part with
section~3.5, which presents our main result, the detection limits of
the global 21-cm signal. We summarize and discuss our conclusions in
section~4.

Hereafter we assume a flat $\Lambda$ CDM cosmology, with matter
density parameter $\Omega_{m}=0.272$ (dark matter plus baryons),
cosmological constant density parameter $\Omega_\Lambda=0.728$,
$H_{0}=70.4 \,{\rm km/s/Mpc}$ (Hubble constant), $\Omega_b=0.045$
(baryons), $n_S=0.963$ (power spectrum index) and $\sigma_8=0.809$
(power spectrum normalization) according to the latest 7-year WMAP
results \citep{jarosik2010}. Unless otherwise stated, we estimate all
errors at the 68.3 percent confidence level.

\section{Modeling the 21-cm signal}

\subsection{The 21-cm foreground and signal model}\label{datdd24}

In general, a global 21-cm measurement yields the antenna temperature
$T_{\rm sky}(\nu)=T_{\rm fg}(\nu)+T_{b}(\nu)$, where $T_{\rm fg}(\nu)$
and $T_{b}(\nu)$ are the foreground and cosmological 21-cm brightness
temperatures, respectively. For the cosmic signal, we assume that the
dipole antenna temperature essentially measures a sky average, since
fluctuations are expected to be present only on angular scales that
correspond to small fractions of the sky. The foregrounds (i.e., our
Galactic emission and radio emission from other galaxies) have
large-scale angular structure, but even if they are convolved with an
angular dipole response, this does not affect our analysis, which only
assumes that they are smooth as a function of frequency.

For the foreground brightness temperature $T_{\rm fg}$, we assume a
polynomial fit of the form
\begin{equation}
\log T_{\rm fg}=\sum_{i=0}^{N_{\rm poly}}a_i \log(\nu/\nu_0)^i\ .
\label{eq:Tb2y}
\end{equation}
In particular, we use the third order polynomial fit from PL, who
fitted the model of the sky put together by \cite{deOliveiraCosta2008}
using all existing observations, by averaging the foregrounds over the
dipole's angular response:
\begin{equation}\begin{split}
\log T_{\rm fg}= \log T_0+a_1\log(\nu/\nu_0) + & a_2[\log(\nu/\nu_0)]^2 \\& 
+a_3 [\log(\nu/\nu_0)]^3\ , 
\end{split}\label{eq:Tb2}
\end{equation}
with parameter values $\nu_0=150{\rm\,MHz}$, $T_0=320{\rm\,K}$,
$a_1=-2.54$, $a_2=-0.074$, and $a_3=0.013$, chosen from fitting to the
band $\nu=100-200$ MHz. Note that at these frequencies $T_{\rm fg}$ is
dominated by diffuse synchrotron radiation from the Galaxy. The
residuals related to such a parameterization of the foreground are
dominated by limitations of the adapted sky model
\citep{deOliveiraCosta2008} and they are $\lesssim1{\,\rm mK}$
averaged over the band. While in principle higher order polynomials
may be needed to reduce such residuals in the future, given the
smoothness of the spectrum of the foreground, low order polynomials
are key to avoid throwing the signal away with the foreground and to
reduce the statistical errors (\S~\ref{datdd455}).


For the cosmic 21-cm signal, the brightness temperature through the
IGM is $T_b=T_{\rm CMB} e^{-\tau}+T_S (1-e^{-\tau})$, with $\tau(z)\ll
1$ the optical depth at $21 (1+z)$ cm produced by a patch of neutral
hydrogen at the mean density and with a uniform 21-cm spin temperature
$T_S$,
\begin{equation} 
\tau(z)\! = \! 9.0 \times 10^{-3} \!\!\left(\!\frac{T_{\rm CMB}} {T_S} \!\right) \!\!\left (
\!\frac{\Omega_b h} {0.03} \!\!\right) \!\!\left(\frac{\Omega_m}{0.3}\right)^{\!\!\!-1/2}\!\!\!
\left(\!\frac{1+z}{10}\!\right)^{\!\!1/2}
\label{eq:Tb34b}
\end{equation}
During the epoch of reionization the Lyman-$\alpha$ and X-ray
radiation backgrounds are expected to be strong enough to bring the
spin temperature $T_S$ to the gas temperature and heat the cosmic gas
well above the cosmic microwave background temperature
\citep{madau1997}. Under these conditions, the observed 21-cm 
brightness temperature $T_b$ relative to the CMB temperature $T_{\rm
CMB}$ is independent of $T_S$. Therefore, $T_b$ (hereafter measured
relative to $T_{\rm CMB}$) is given by:
\begin{equation}\begin{split}
\!\!\!T_b(z)=& (T_S-T_{\rm CMB}) (1-e^{-\tau}) Q_{\rm H\:I} \! 
 =\\  = & T_{21} \left( \frac{1+z} {10} \right)^{1/2} Q_{\rm H\:I} 
\end{split}\label{eq:Tb34}
\end{equation}
where $T_{21}= 9.0 \times 10^{-3}(\Omega_b h/0.03)
(\Omega_m/0.3)^{\!-1/2}\,T_{\rm CMB} \simeq 27.2{\,\rm mK}$, and
$Q_{\rm H\:I}={N_{\rm H\:I}}/{(N_{\rm H\:I}+N_{\rm H\:II})}$ is the
neutral hydrogen fraction. Note that the ionized fraction is $Q_{\rm
H\:II}=1-Q_{\rm H\:I}$. Throughout this paper, given that we are
interested in the spatially-integrated 21-cm signal, we consider only
the cosmic mean neutral or ionized fraction, and neglect spatial
fluctuations in the 21-cm signal from density and peculiar velocity
fluctuations.

We consider an experiment covering the frequency range $100-250{\,\rm
MHz}$ in 50 bins of bandwidth $B=3$ MHz for each of the receiver
frequency channels, and integrating time $t_{\rm int}=500$ hours (these
parameters mimic EDGES with an order of magnitude longer integration
time). Under these assumptions, the thermal noise in the $i$'th
receiver frequency channel is given by the radiometer equation:
\begin{equation}
\sigma_i^2=\frac{T_{\rm sky}^2(\nu_i)}{B t_{\rm int}},
\label{eq:Tb34ee}
\end{equation}
We note that the frequency range we consider corresponds to the
redshift range 4.7--13.2. 

Our model thus consists of the foreground brightness temperature
$T_{\rm fg}(\nu)$ and a suitable model for the cosmological 21-cm
signal $T_{b}(\nu)$. To derive the parameter errors, we directly
calculate the Fisher matrix of the foreground plus 21-cm signal
parameters ${\bf p}$ expected with the above thermal noise $\sigma_i$,
\begin{equation}
F_{ij}=\sum_{n=1}^{N_{\rm
channel}}\frac{1}{\sigma_n^2}\left(\frac{\partial T_{\rm
sky}(\nu_n;{\bf p})}{\partial p_i}\frac{ \partial T_{\rm
sky}(\nu_n;{\bf p})}{\partial p_j}\right)\ .
\label{fisher1}
\end{equation}
This equation provides an estimate of the covariance matrix
$C=F^{-1}$, and therefore of the parameter uncertainty in dipole
observations. Note that this is equivalent to finding the covariance
matrix near the minimum $\chi^2$. These errors should be accurate as
long as they are small. However, in many cases we consider regions of
parameter space where the errors are large, e.g., when we calculate
the detection limit of an experiment, or more generally due to
parameter degeneracies. Thus, we often use a more generally-valid and
computationally intensive Monte-Carlo (MC) error analysis. We generate
a large number of MC simulations of the measurement noise, finding the
best-fit parameters in each case by minimizing the $\chi^2$:
\begin{equation}
\chi^2=\sum_{n=1}^{N_{\rm
channel}}\frac{1}{\sigma_n^2}\left[\Delta T_{\rm
sky}(\nu_n;{\bf p})\right]^2\ ,
\label{chi2}
\end{equation}
where $\Delta T_{\rm sky}$ is the difference between the measured and
predicted total 21-cm sky temperature in channel $n$ (centered at the
frequency $\nu_n$). The distribution of best-fit parameters in the MC
trials yields parameter errors and their correlations.

In the following sections we will focus on the modeling of the 21-cm
signal, in particular of the neutral fraction $Q_{\rm H\:I}$ in
equation~(\ref{eq:Tb34}). Given the great uncertainty associated the
evolution of the neutral fraction $Q_{\rm H\:I}$ due to the uncertain
astrophysics of the ionizing sources, we will begin with a toy model,
namely the {\em tanh}-based parameterization used by previous authors,
which simply characterizes when and for how long the reionization
occurs; then we will consider a more complex and physically motivated
structure formation model, in order to better describe the
reionization process and extract interesting astrophysical
information, such as the mass of the smallest galaxies that can form
and contribute to the IGM ionization, the overall number of
ionizations per baryon and the redshift evolution of the ionizing
sources.


\subsection{The {\em tanh}-based model of reionization} \label{sec:reion}

The {\em tanh}-based parameterization is characterized by two
parameters describing the two main features of reionization: its
mid-point $z_r$ and duration $\Delta z$. This approach was used by PL
for the 21-cm signal (note that \cite{bowman2008} used a somewhat
different parameterization), and a similar {\em tanh}-based fitting
function is the default parameterization of reionization in CAMB
(although there it is based on the optical depth for CMB scattering)
\citep[][]{lewis2008}. Under the assumptions outlined above for
the gas state during reionization, the 21-cm signal is given by
\begin{equation}\label{tanh}
T_{b}(z)=T_{21}\left(\frac{1+z}{10}\right)^{1/2}
\frac{1}{2}\left[\tanh\left(\frac{z-z_r}{\Delta z}\right)+1\right].
\end{equation}
Note that $z_r$ is the redshift at which the ionized fraction $Q_{\rm
H\:II}=50\%$, while $z_r+\Delta z$ and $z_r-\Delta z$ are the
redshifts at which $Q_{\rm H\:II}=11.9\%$ and 88.1\%,
respectively. This parameterization is a convenient mathematical toy
model but it does not have any particular physical motivation. We
consider both the case where we fix the amplitude of the signal
$T_{21}$ to its known value (equation~\ref{eq:Tb34}), and the case
where we leave it as a free parameter (following PL).

\subsection{A simple CDM-dominated galaxy formation model}\label{datdd2}

In the previous section we considered a toy model that has been used
in previous observational and theoretical papers. While a toy model
can be justified as an unbiased analysis tool, especially given the
large current uncertainty in the astrophysics of high-redshift
galaxies, such an approach is also problematic. The particular model
assumed (with a fixed, arbitrarily-chosen shape) may lead to
systematically biased results if it cannot reasonably approximate the
real reionization (we consider this issue further below). In addition,
it can be hard to interpret any results of a toy model in terms of the
underlying parameters of interest. In particular, the redshift
evolution of reionization is closely related to structure
formation. Indeed, reionization is driven by the intergalactic
ionizing radiation field, which (we expect) is the result of the
ionizing radiation escaping from stars and quasars within galaxies.
While astrophysical aspects (such as star formation and feedback) play
a significant role, the evolution of galaxies is driven by the
properties of the host dark matter halos. A major reason for studying
reionization is to learn more about both galaxy formation and the
astrophysical properties of galaxies in the reionization era. Thus, a
more realistic and useful approach is to use models based on our
understanding of CDM-driven galaxy and structure formation, a model
with many successes at lower redshifts, and to include some
flexibility in order to account for the uncertain astrophysical
parameters. Here we take the first step in this process by using a
simple model that is based on the standard theory of galaxy formation.

We begin with the equation from \citet{barkana2001}, based on
\citet{SG87}, that statistically describes the transition from a
neutral universe to a fully ionized one; in particular this equation
describes the evolution of the ${\rm H\:II}$ filling factor $Q_{\rm
H\:II}$, i.e., the fraction of the volume of the universe which is
filled by ${\rm H\:II}$ regions.
\begin{equation}
\frac{dQ_{\rm H\:II}}{dt}=\frac{N_{\rm ion}}{0.76} \frac{dF_{\rm
col}}{dt}- \alpha_B \frac{C}{a^3} {\bar{n}}_H^0 Q_{\rm H\:II}
\label{QIIeqn}\ , 
\end{equation}
assuming a primordial mass fraction of hydrogen of 0.76. In this
equation $N_{\rm ion}\equiv N_{\gamma} \, f_{\rm star}\, f_{\rm esc}$
is an efficiency parameter that gives the overall number of
ionizing photons per baryon; for instance, if we assume that baryons are
incorporated into stars with an efficiency of $f_{\rm star}=10\%$, the
escape fraction for the resulting ionizing radiation is $f_{\rm
esc}=5\%$ and $N_{\gamma}\approx 4000$ ionizing photons are produced
per baryon in stars (for a stellar IMF similar to the one measured
locally but with a metallicity equal to $1/20$ of the solar value), we
infer that for every baryon in galaxies $\sim$20 escaping ionizing
photons are produced by stars. We obtain a similar result if we
consider mini-quasars rather than stars \citep{barkana2001}. It is
possible to get a substantially higher $N_{\rm ion}$ using Pop III
stars or by assuming a high escape fraction. $N_{\rm ion}$ also
determines the maximum comoving radius of the region that a halo of
mass $M$ can ionize on its own (neglecting recombinations),
\begin{equation}
r_{\rm max}=
675\, {\rm kpc} \left( \frac{N_{\rm ion}}{40}\,
\frac{M} {10^9 M_{\sun}}\right)^{1/3}\ , \label{rmax} 
\end{equation}
a radius that is larger than the halo virial radius by a factor of
$\sim 20$ (essentially independent of redshift and halo mass).

Also in equation~(\ref{QIIeqn}), $a=1/(1+z)$ is the scale factor,
${\bar{n}}_H^0$ is the present number density of hydrogen, $\alpha_B$
is the case B recombination coefficient of hydrogen, and $C$
represents the volume-averaged clumping factor (in general
time-dependent),
\begin{equation} 
C=\left<n_H^2\right>/{\bar{n}}_H^2\ . \label{clump}
\end{equation}
This factor crudely accounts for a non-uniform IGM that includes
high-density clumps. Since each ionized bubble is far larger than the
typical scale of clumping, so that many clumps are averaged over, $C$
can be assumed to be approximately spatially uniform.

The collapsed fraction $F_{\rm col}$ is the fraction of all the
baryons in the universe that is in galaxies, i.e., the fraction of gas
which settles into halos and cools efficiently inside them. A simple
estimate of the collapse fraction at high redshift is the halo mass
fraction above some cooling threshold. More generally, we include
halos above some minimum circular velocity $V_c$. We use the
Sheth-Tormen halo mass function, which accurately fits the mean halo
abundance in simulations \citep{sheth2002}. We calculate the power
spectrum transfer function using the {\it CAMB} code
\citep{lewis2000}.

The solution of equation~(\ref{QIIeqn}) is \citep{barkana2001}
\begin{equation}  Q_{\rm H\:II}(t) =\int_{0}^t
\frac{N_{\rm ion}} {0.76} \frac{dF_{\rm col}}{dt'}\,e^{F(t',t)} dt'\ , 
\label{sol1}
\end{equation} 
where (if $C$ is time-independent) 
\begin{equation}
F(t',t) = -\frac{2}{3}\frac{ \alpha_B
{\bar{n}}_H^0} {\sqrt{{\Omega_m}} H_0}\,C \left[f(t')-f(t)\right]\ ,
\end{equation}
and where (in flat $\Lambda$CDM)
\begin{equation} 
f(t)=\sqrt{\frac{1}{a^3}+\frac{1-{\Omega_m}}{{\Omega_m}}}\ . \label{fLCDM} 
\end{equation} 
Once $Q_{\rm H\:II}(t)$ reaches unity, the universe becomes fully
reionized and remains so within our model. 

Equation~(\ref{sol1}) allows us to quickly calculate the time
evolution of the ionized fraction of the universe once we fix the IGM
clumping factor and the parameters related to the ionizing
sources. Also, in equation~(\ref{fisher1}) we calculate accurate
derivatives for the 21-cm signal as numerical integrals of partial
derivatives of the integrand in equation~(\ref{sol1}). Hereafter, we
refer to this CDM-dominated galaxy formation model as the CDM model.

Within the CDM model, the parameters that determine the redshift
evolution of $Q_{\rm H\:II}$ (and $T_b$) are $N_{\rm ion}$ (which we
assume is a constant, in this first investigation of fitting global
21-cm signals from a galaxy formation model), C (likewise assumed
constant), and the minimum halo circular velocity $V_c$ (equivalent to
a minimum mass) required for halos that host galaxies. We allow $V_c$
to vary, since while cooling sets a minimum value for it, feedback
(radiative or from supernovae) may in reality set a higher threshold
for effective star formation. We set $C=1$ as our standard value
(i.e., corresponding to a uniform IGM), and discuss in several cases
the effect of allowing $C$ to vary.

\section{Results}

\subsection{Global properties of the two models}

\begin{figure}
\begin{center}
\psfig{figure=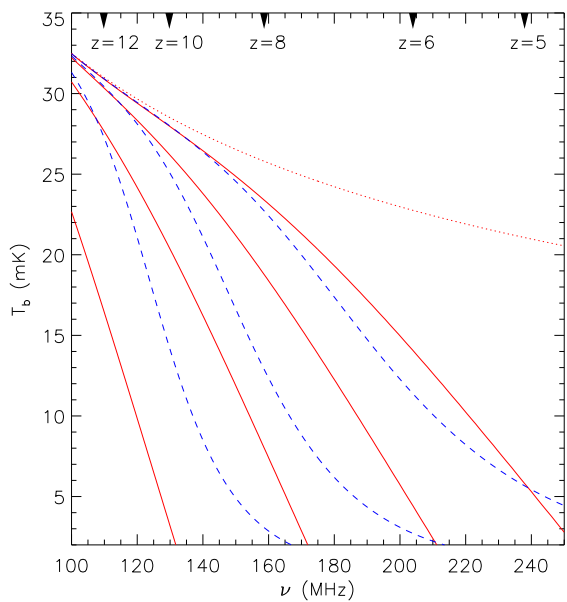,width=0.48\textwidth}
\caption[]{Global 21-cm signal predicted by the CDM model (solid curves) 
for various values of the minimum circular velocity of galactic halos
($V_c=\{4.5, 16.5, 36, 64\}$ from left to right). We fix $C=1$ and
$N_{\rm ion}=20$. We also show (dashed curves) the global 21-cm signal
for a {\em tanh} model of reionization with $\Delta z=2$ and
$z_r=\{6,8,10\}$ (from right to left). The signal for a fully neutral
universe is shown for comparison (dotted curve). Note some redshift
values (at the top) that fall within the experimental frequency range
($z=4.7-13.2$).}
\label{entps3xc}
\end{center}
\end{figure}

We begin by visually comparing our two models, the {\em tanh}-based
toy model and the more realistic CDM-based model. In
Figure~\ref{entps3xc} we plot a few examples of the global
(volume-averaged) 21-cm signal for each model, over the assumed
experimental frequency range (100--250 MHz). For the CDM model, we
consider a minimum halo circular velocity $V_c=\{4.5, 16.5, 36, 64\}$
km/s, corresponding to a midpoint of reionization $z_r=\{11.8, 8.7,
6.9, 5.6\}$ and a minimum galactic halo of $1.3\times10^6,
1.0\times10^8, 1.4\times10^{9}, 1.0\times10^{10}\; M_{\odot}$ at
$z=z_r$, with the other parameters fixed at $N_{\rm ion}=20$ and
$C=1$. The two lower values of $V_c$ correspond to cooling via
molecular hydrogen and via atomic hydrogen and helium, respectively,
while the higher values are roughly in the range of values possible
due to photoheating or supernova feedback. In general, a higher
circular velocity implies that only more massive halos are included,
delaying reionization to a lower redshift (for a fixed $N_{\rm
ion}$). In addition, since more massive halos are more rare and
correspond to the Gaussian tail of large (positive) density
fluctuations, their abundance changes rapidly with time, thus making
reionization more rapid in terms of its redshift extent; this would be
obvious in a comparison of models all normalized to the same $z_r$,
while the effect is suppressed in this figure since the lower
redshifts are stretched into a relatively large frequency
interval. The main point is that $z_r$ is determined by a combination
of $V_c$ and $N_{\rm ion}$, while the extent of reionization is
separately sensitive to $V_c$, so that a measurement of the
reionization history can probe both the characteristic halo mass of
the ionizing sources and their ionizing efficiency.

The main qualitative difference between the two models is that the CDM
model shows a steady rise of $Q_{\rm H\:II}$, while the toy model is
much more round in shape, in particular showing a slowdown of
reionization during its last quarter or so. The toy model is
explicitly symmetric in redshift about the midpoint of reionization,
while in the CDM model reionization starts slowly but ends
quickly. The steady acceleration of reionization in the CDM model is
driven by the exponential rise of the ionizing sources, which
correspond to rare halos at these redshifts. However, our simple model
is by no means fully general, so we treat our conclusions with
caution, as discussed further below. We consider the CDM model to be
an example of a realistic model, which may be quantitatively plausible
if some of the missing complications turn out to have a relatively
minor effect on the global 21-cm signal. We note,
however, that some complications may tend to make the CDM model more
similar in shape to the tanh model (see Sect. 4).

The {\em tanh}-based model is explicitly expressed in terms of the
mid-point $z_r$ and duration $\Delta z$ of reionization, while in the
CDM model these are derived parameters. While the midpoint $z_r$ is
naturally defined as $Q_{\rm H\:II}=50\%$, there is some ambiguity in
$\Delta z$. For the toy model, we have chosen to follow previous PL in
defining $\Delta z$ as above, a definition that is natural for the
{\em tanh} function, and implies that $z_r+\Delta z$ and $z_r-\Delta
z$ (i.e., a total spread of $2 \Delta z$) delineate the central 76.2\%
of reionization. However, for the CDM model we use a definition that
should be the natural one more generally: $\Delta
z=(z_{-1\sigma}-z_{+1\sigma})/2$, with $z_{-1\sigma}$ and
$z_{+1\sigma}$ being the redshifts corresponding to $Q_{\rm
H\:II}=0.16$ and $Q_{\rm H\:II}=0.84$, respectively. Thus, in the CDM
model a spread of $2 \Delta z$ marks the central $68\%$. In the
context of the {\em tanh} model, this definition would give a value of
$\Delta z$ smaller by a factor of 1.2 than the definition that we have
followed.

\begin{figure}
\begin{center}
\psfig{figure=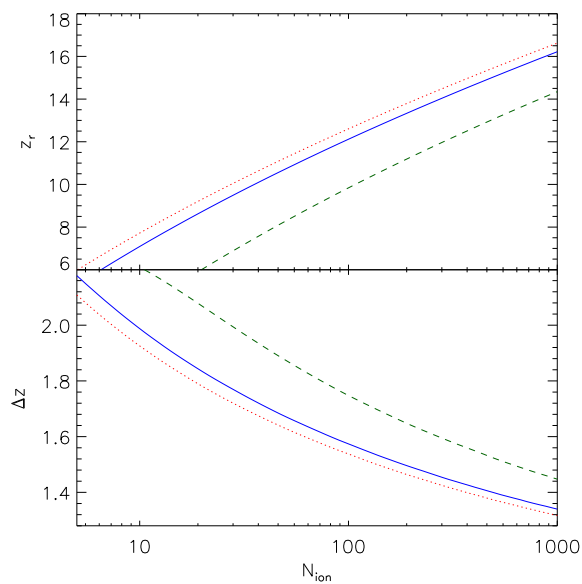,width=0.48\textwidth}
\caption[]{Reionization characteristics $z_r$ and $\Delta z$ as 
a function of $N_{\rm ion}$ for the CDM model. We fix $V_c=16.5$ km/s
and consider $C=\{0,1,10\}$ (dotted, solid and dashed curves,
respectively).}
\label{ent246e}
\end{center}
\end{figure}

In order to gain intuition on how the characteristics of reionization
are set in the CDM model, in Figure~\ref{ent246e} we show the
dependence of $z_r$ and $\Delta z$ on $N_{\rm ion}$ for one value of
$V_c$ and several values of $C$. Larger values of $N_{\rm ion}$ lead
to earlier reionization (i.e., higher $z_r$) at a time when the
ionizing sources are brighter and rarer, so their rarity leads to a
shorter span $\Delta z$ for reionization. We compare $C=1$ to no
recombinations $C=0$ and fast recombinations ($C=10$). At least during
most of reionization $C$ is likely to be of order unity, since the
low-density IGM gets reionized first, and the denser gas is left for
the final stages of reionization. We find that a high clumping factor
can be essentially counterbalanced by a higher value of $N_{\rm ion}$,
at least during the central portion of reionization that defines $z_r$
and $\Delta z$, so that including $C$ as a free parameter mostly
includes the degeneracy of the parameters but does not significantly
change the allowed parameter space of $z_r$ and $\Delta z$. 

\begin{figure}
\begin{center}
\psfig{figure=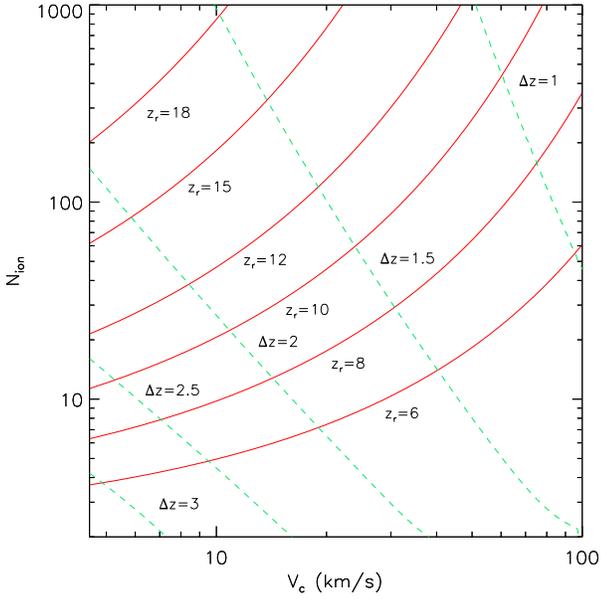,width=0.48\textwidth}
\caption[]{Isocontours of $z_r$ (solid lines) and $\Delta z$ (dashed 
lines) derived from the CDM model in the $N_{\rm ion}$-$V_c$ plane
(for $C=1$). The parameters ($N_{\rm ion},V_{\rm c}$) are allowed to vary over $N_{\rm ion}\in \{2,1000\}$ and $V_{\rm c}\in \{4.5,100\}$ km/s.}
\label{ent2rt2}
\end{center}
\end{figure}

A more complete picture of the allowed parameter space is shown in
Figure~\ref{ent2rt2}, where we present the isocontours of $z_r$ and
$\Delta z$ in the $N_{\rm ion}$-$V_c$ plane. For reasonable values of
$N_{\rm ion}$ and $V_c$ the reionization midpoint $z_r$ varies widely
(from below 6 to above 18), while the span of reionization $\Delta z$
covers roughly the range 1--3. As noted above, these values of $\Delta
z$ should be multiplied by a factor of 1.2 for a fair comparison with
$\Delta z$ in the {\em tanh} model.

\subsection{Expected parameter errors: the {\em tanh} model}

\begin{figure}
\begin{center}
\psfig{figure=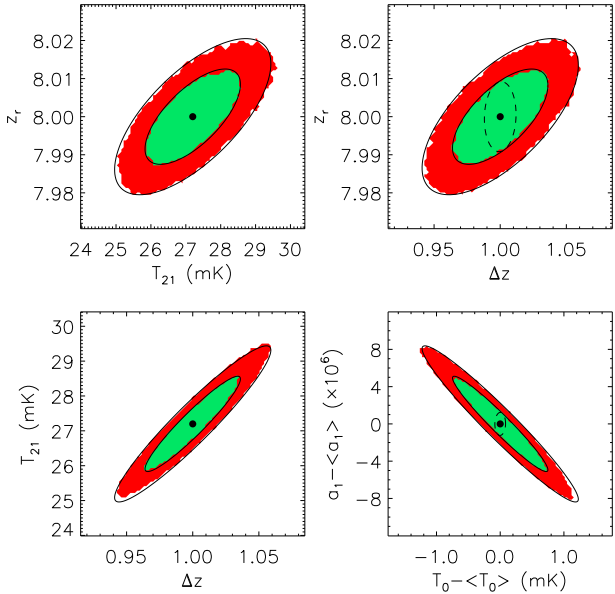,width=0.48\textwidth}
\caption[]{The 68 and 95\% confidence regions of various
parameter pairs for the {\em tanh} model of reionization, comparing
the MC likelihood (green/bright and red/dark shaded regions,
respectively) to the Fisher matrix (solid ellipses) calculations. For
comparison with PL, we set $z_r=8$ and $\Delta z=1$, $T_{21}$ free and
fit four polynomial (foreground) parameters (i.e., $N_{\rm
poly}=3$). In the panels on the right, we also plot the Fisher matrix
(dashed ellipses) results for the same model with $T_{21}$ fixed at
its known value. We assume an integration time $t_{\rm int}=500$
hours.}
\label{ent2rr}
\end{center}
\end{figure}

We now derive parameter errors for some specific instances of
potential global 21-cm observations. We begin with the {\em tanh}
model, and test the Fisher matrix formalism against the MC error
analysis. Taking fiducial values of $z_r=8$, $\Delta z=1$, and
assuming the model parameters for the foreground as in
\S~\ref{datdd24}, we generate $10^5$ MC
simulations of the noise, finding the best fit parameters in each
case. The resulting parameter contours are shown in
Figure~\ref{ent2rr} along with the corresponding Fisher matrix
constraints. We find good agreement between the two methods in this
example.

We consider both the case where $T_{21}$ is a free parameter (as
assumed by PL) and where it is fixed at its known value.  The error
ellipses show that there is a strong positive correlation between
$T_{21}$ and $\Delta z$, i.e., there is an uncertainty in
distinguishing between a higher amplitude extended scenario and a
lower amplitude quicker scenario, since both produce a similar slope
with frequency in the 21-cm signal, and it is this sharp slope that
can be distinguished from the foregrounds (which are smooth and thus
can be modeled by a low-order polynomial). There are also significant
correlations among the other parameters. Comparing with PL, we note
that the amplitude of the errors that we find are smaller in all four
panels, and also the sign of the correlation is different in the
$z_r-T_{21}$ and $z_r-\Delta z$ relations. Note that PL used a different frequency interval, i.e. $100-200{\,\rm MHz}$ (Pritchard \& Loeb, personal communication). 

The results thus show significant correlations among the parameters
when $T_{21}$ is free, but substantially reduced errors and
correlations when $T_{21}$ is fixed. We conclude that it is possible
to obtain a direct observational estimate for $T_{21}$ from these type
of data, in order to check consistency with the theoretically expected
value, but in order to constrain reionization it is very helpful to
use our independent knowledge of $T_{21}$. In this example we have
assumed an integration time $t_{\rm int}=500$ and fitted a foreground
polynomial of degree $N_{\rm poly}=3$ over the entire frequency range,
assuming no remaining foreground or systematic residuals. This
represents a quite optimistic assumption regarding the level of
systematic noise and the ease of foreground removal, far beyond the
current EDGES experiment, as discussed further below.

\begin{figure}
\begin{center}
\psfig{figure=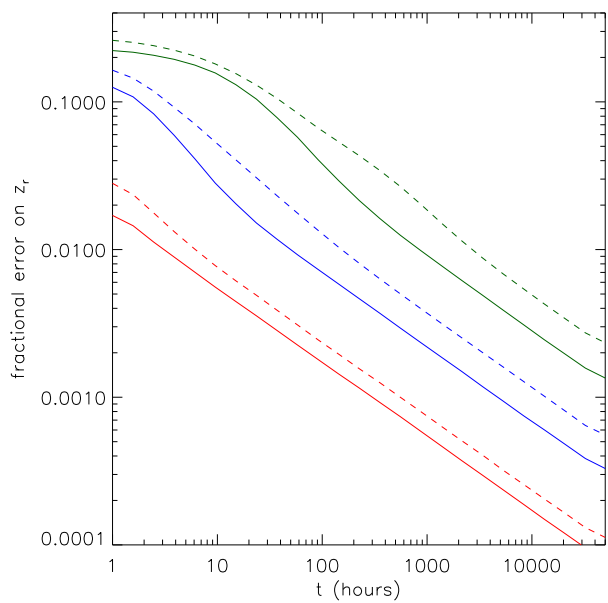,width=0.48\textwidth}
\caption[]{Relative error on $z_r$ as a function of $t_{\rm int}$ 
for $\Delta z=\{1,2,3\}$ (from bottom to top in each case) and $z_r=8$
for the {\em tanh} model of reionization. We consider $T_{21}$
fixed or free in the analysis (solid and dashed curves,
respectively). The errors have been calculated via MC analysis.}
\label{ent245}
\end{center}
\end{figure}

Since we have just considered a rather optimistic experimental
scenario, it is interesting to consider more realistic possibilities.
One way to do this is to vary the integration time, thus increasing
the errors. We can take this also as a rough indication of the effect
of increasing the foreground or systematic residuals to various levels
(still with $N_{\rm poly}=3$). In the case considered in
Figure~\ref{ent2rr}, the errors per frequency bin range from 0.4 mK in
the lowest-frequency bin to an order of magnitude lower at the
highest-frequency bin. More generally, the noise varies with the
integration time $t_{\rm int}$ of the bolometer as $t_{\rm
int}^{-0.5}$ [equation~(\ref{eq:Tb34ee})]. In Figure~\ref{ent245} we
consider the fractional error on $z_r$ as a function of $t_{\rm int}$
for $\Delta z=\{1,2,3\}$ and $z_r=8$. Note that the fractional error
in $z_r$ varies approximately as $t_{\rm int}^{-0.5}$ since the errors are
relatively small over most of the plotted range (which makes the model
behave approximately like a linear model). More extended reionization
scenarios increase the errors significantly. Fixing $T_{21}$ at its
known value reduces the errors by $15-50\%$.

We will directly consider detection limits in a later section, but one
way to define a successful detection of reionization is when
observations yield a meaningful constraint on the most interesting
single number associated with reionization, namely $z_r$. Within the
{\em tanh} model, rough ($10\%$) constraints on $z_r$ are
expected for $t_{\rm int}=26$ hours (if $\Delta z=3$) or 1.9 hours (if
$\Delta z=2$), while tight ($1\%$) constraints require $t_{\rm
int}=848$ hours (if $\Delta z=3$), 51 hours (if $\Delta z=2$), or 3.1
hours (for sharper reionization, with $\Delta z=1$).

\begin{figure}
\begin{center}
\psfig{figure=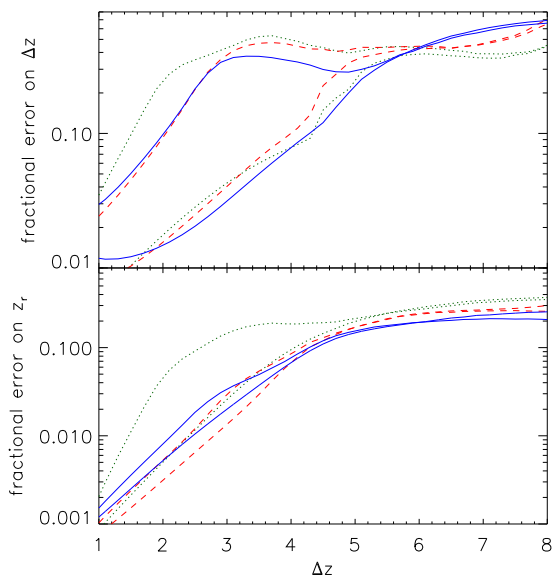,width=0.48\textwidth}
\caption[]{Relative error on $\Delta z$ and $z_r$ as a function of 
$\Delta z$ for the {\em tanh} model of reionization, for
$z_r=\{6,8,10\}$ (dotted, dashed and solid curves, respectively).  In
each case we consider $T_{21}$ to be fixed or free (where fixed
corresponds to the lower curve at the left end of the plot). The
errors have been calculated via MC analysis.}
\label{ent246}
\end{center}
\end{figure}

In Figure~\ref{ent246} we show a different range of the parameter
space, considering three possible values of $z_r$ while varying
$\Delta z$ over a wide range, all for $t_{\rm int}=500$ hours. Here we
show the relative errors on both $\Delta z$ and $z_r$, finding that
$z_r$ is generally better constrained, by up to an order of
magnitude. The errors increase with $\Delta z$, roughly saturating at
30\% for $z_r$ and 50\% for $\Delta z$ (i.e., the errors only increase
slowly beyond these values as $\Delta z$ is further increased beyond
$\sim 5$). As before, fixing $T_{21}$ at its known value can make a
big difference (compared to allowing it to be a free parameter),
especially in constraining $\Delta z$ (except when all the errors are
large, for high $\Delta z$). The relative errors vary weakly with
$z_r$ over the range of 6--10 (a range which is all well within our
assumed experimental frequency window).

\subsection{Expected parameter errors: the CDM model}

\begin{figure}
\begin{center}
\psfig{figure=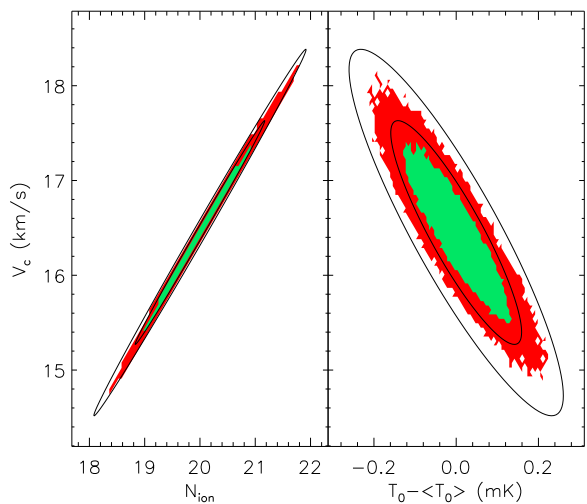,width=0.48\textwidth}
\caption[]{The 68 and 95\% confidence regions of various
parameter pairs for the CDM reionization model, comparing the MC
likelihood (green/bright and red/dark shaded regions, respectively) to
the Fisher matrix (solid ellipses) calculations. We set our model
parameters $(N_{\rm ion}, C, V_c)$ to the fiducial values of
$(20,1,16.5)$ and assume an integration time $t_{\rm int}=500$ hours
and a foreground (plus systematics) polynomial with $N_{\rm poly}=3$.}
\label{ent2}
\end{center}
\end{figure}

We begin our exploration of global 21-cm measurements in the context
of the CDM model with Figure~\ref{ent2}, where we show parameter
errors and correlations for the fiducial values of $(20,1,16.5)$ for
$(N_{\rm ion}, C, V_c)$, respectively. The error ellipses show that
there is a strong positive correlation between $V_c$ and $N_{\rm
ion}$; indeed, this is a partial degeneracy, since while the error
ellipse covers a small total area, each of these parameters is
uncertain at a relatively high ($\sim 10\%$) level. From
Figure~\ref{ent2rt2} it is apparent that this degeneracy with a
positive correlation is driven by the value of $z_r$, which is the
main constraint from these observations (at least in the example we
are considering of a high-precision experiment with low noise). There
is also a strong anti-correlation between $V_c$ and
$T_0-\left<T_0\right>$, demonstrating how the foreground fitting
removes power from the total signal $T_{\rm sky}(\nu)$ making it more
difficult to determine the parameters of reionization. The Fisher
matrix and MC formalisms yield a reasonable agreement, but there are
bigger differences compared to the {\em tanh} model, likely because
the partial degeneracy gives larger errors in some directions in the
CDM model.

\begin{figure}
\begin{center}
\psfig{figure=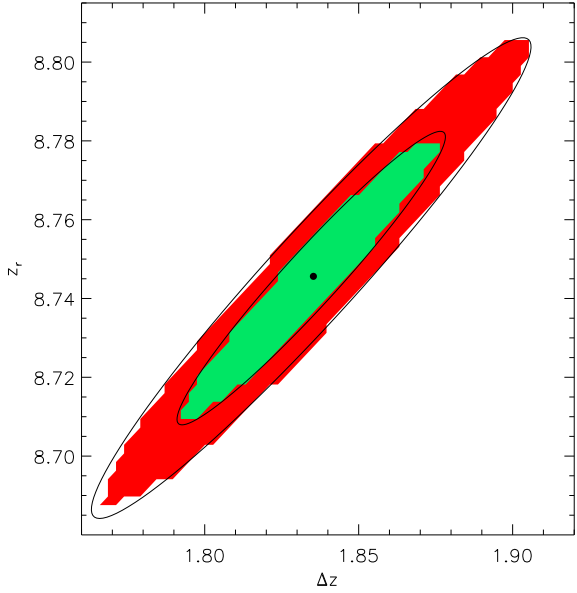,width=0.48\textwidth}
\caption[]{The 68 and 95\% confidence region of the 
reionization characteristics $z_r$ and $\Delta z$ derived from the CDM
model, comparing the MC likelihood (green/bright and red/dark shaded
regions, respectively) to the Fisher matrix (solid ellipses)
calculations.  We set our model parameters $(N_{\rm ion}, C, V_c)$ to
the fiducial values of $(20,1,16.5)$ and assume $t_{\rm int}=500$
hours and $N_{\rm poly}=3$. The black dot corresponds to the input
model values $(z_r,\Delta z)=(8.74,1.83)$.}
\label{ent2rt}
\end{center}
\end{figure}

Fortunately, the partial degeneracy in the parameters of the CDM model
is relatively harmless in terms of measuring the characteristics of
reionization. For the case considered in Figure~\ref{ent2}, we
measure $z_r=8.74\pm0.02$ and $\Delta z=1.83\pm0.02$. The
two-parameter contour is shown in Figure \ref{ent2rt}. The relative
errors in $z_r$ and $\Delta z$ are much smaller than in $V_c$ and
$N_{\rm ion}$, showing that the global 21-cm measurements constrain
these quantities rather directly, somewhat independently of the
underlying galaxy and halo parameters. Also, as noted above, the
fractional error on $z_r$ is significantly smaller than in $\Delta z$.
The plotted results assume $C=1$, but we find that if we allow the
clumping factor to be a free parameter in the fit, this increases the
CDM model parameter degeneracies but it does not significantly affect
the errors on $z_r$ and $\Delta z$.

\begin{figure}
\begin{center}
\psfig{figure=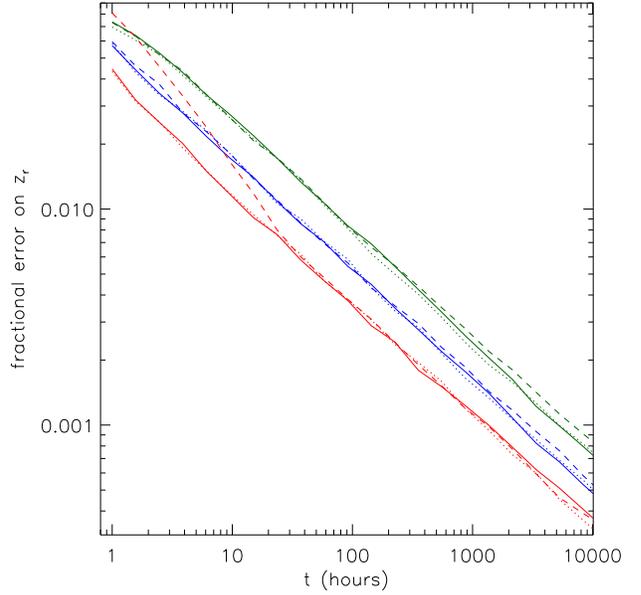,width=0.48\textwidth}
\caption[]{Relative error on $z_r$ as a function of $t_{\rm int}$ 
for a CDM model of reionization with $\Delta z=\{1.5,2,2.5\}$ (black,
red and blue colors, respectively; also from bottom to top in each
case). We set $z_r=8$ and consider $C=0$, $C=1$ and $C=10$ (dotted,
solid and dashed curves, respectively). The errors have been
calculated via MC analysis, with $C$ held fixed in the fitting.}
\label{ent268}
\end{center}
\end{figure}

As we did for the {\em tanh} model, we vary the integration time and
consider the expected experimental accuracy in measuring the most
important quantity, $z_r$. Figure~\ref{ent268} shows the fractional
error on $z_r$ for $\Delta z=\{1.5,2,2.5\}$. As in
Figure~\ref{ent245}, the fractional error varies approximately as
$t_{\rm int}^{-0.5}$, and increases for larger $\Delta z$. For $C=1$,
tight ($1\%$) constraints on $z_r$ require $t_{\rm int}=68$ hours (if
$\Delta z=2.5$), 29 hours (if $\Delta z=2$), or 13 hours (if $\Delta
z=1.5$). The CDM model gives somewhat better accuracy than the {\em
tanh} model, for similar values of $z_r$ and $\Delta z$, though the
numbers are comparable.

\subsection{Systematic effect of the choice of reionization 
model}\label{datdd454}

\begin{figure}
\begin{center}
\psfig{figure=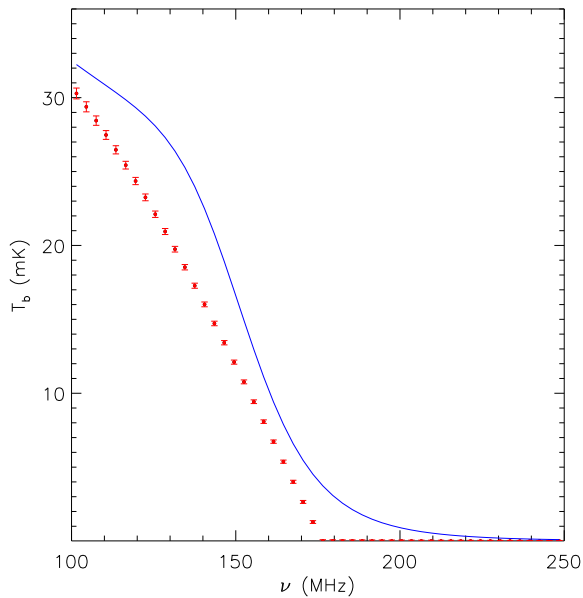,width=0.48\textwidth}
\caption[]{Fit of a {\em tanh} model + foreground to 21-cm 
data generated from a CDM model + foreground. We show just the 21-cm
signal (without the foreground), for the input CDM model (points with
error bars showing the measurement errors) and for the best-fit {\em
tanh} model (solid curve). The input model sets $(N_{\rm ion}, C,
V_c)$ to the fiducial values of $(20,1,16.5)$. We obtain
$z_r=8.19\pm0.01$ and $\Delta z=1.28\pm0.01$ as best fit parameters
with $\chi^2=1044$ (for 44 degrees of freedom), while the true values
for the CDM model are $z_r=8.74$ and $\Delta z=1.83$. We assume
$t_{\rm int}=500$ hours and $N_{\rm poly}=3$.}
\label{zahn2007b}
\end{center}
\end{figure}

Our use of two different models allows us to explore the systematic
effects of assuming an incorrect model when trying to reconstruct the
global 21-cm signal from observations. We assume our more realistic
CDM model as the input model, and try to fit the resulting 21-cm
signal with the {\em tanh} model. In Figure~\ref{zahn2007b} we plot
the 21-cm signal as inferred from the fit of the {\em tanh} model +
foreground to the 21-cm data generated from the CDM model + foreground
with $(N_{\rm ion}, C, V_c)=(20,1,16.5)$, corresponding to $z_r=8.74$
and $\Delta z=1.83$. The fit of the {\em tanh}-based model +
foregrounds, after the subtraction of the best-fit foreground
polynomial (which takes out part of the signal together with the
original input foreground), leads to a quite different output profile
of $T_{b}(\nu)$ compared to the input one, and to biased values of the
midpoint and duration of reionization. The best-fit parameters are
$z_r=8.19\pm0.01$ and $\Delta z=1.28\pm0.01$; the latter is even more
discrepant than may appear, because the input CDM value of $\Delta z$
should be multiplied by 1.2 for a fair comparison with the {\em tanh}
model. While the statistical errors of the fit are tiny (for $t_{\rm
int}=500$ hours and $N_{\rm poly}=3$), the systematic errors are quite
large. The systematic errors are related both to the inadequacy of the
{\em tanh} model in representing the reionization signal and to the
presence of the foreground which must be fit with a polynomial.

The news, though, is not all bad, since an experiment with such low
noise levels would result in high, strongly discrepant, $\chi^2$
values for such a poor fit, giving a clear indication that the
template being used must, indeed, be modified. In particular, we find
$\chi^2=1044$ for 44 degrees of freedom. This means that in this
example, only a much reduced experimental sensitivity corresponding to
$t_{\rm int} \sim 20$ hours would give a reduced $\chi^2$ of order
unity for the {\em tanh} model fit. Any experiment above this
sensitivity would be able to discriminate between the CDM and {\em
tanh} models.

\begin{figure}
\begin{center}
\psfig{figure=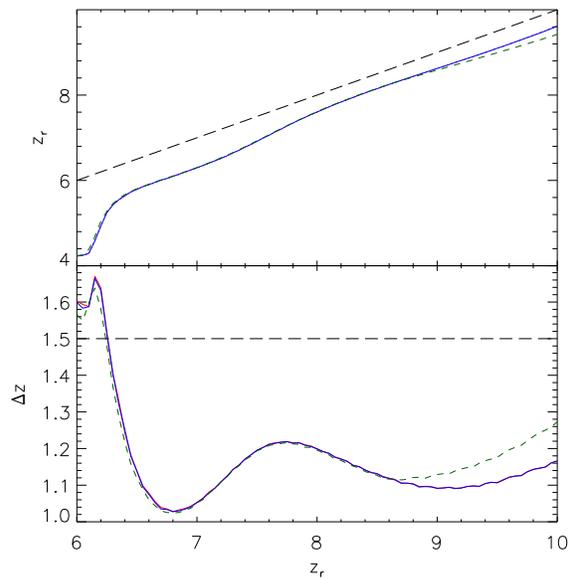,width=0.48\textwidth}
\caption[]{Values of $z_r$ and $\Delta z$ from the fit of a 
{\em tanh} model + foreground to 21-cm data generated from a CDM model
of reionization with $\Delta z=1.5$, as a function of the input $z_r$
of the CDM model. We consider $C=0$, $C=1$ and $C=10$ (dotted, solid
and short-dashed curves, respectively); the $C=0$ and $C=1$ cases are
indistinguishable. We also indicate the input values for the CDM model
(long-dashed curves). $T_{21}$ has been kept fixed to its known value
in the fits. We have assumed $N_{\rm poly}=3$. This plot is independent
of $t_{\rm int}$.}
\label{zahn2007c}
\end{center}
\end{figure}

In Figure~\ref{zahn2007c} we explore these kind of systematic errors
over a wider range of the parameter space. We compare the best-fit
parameters $z_r$ and $\Delta z$ from fitting the {\em tanh} model +
foreground with the true values for an input CDM model of
reionization.  We fix the input $\Delta z=1.5$ (equivalent to $\sim
1.8$ in the {\em tanh} model), and vary $z_r$ over the range
6--10. This wider range shows similar results to the example above,
where the best-fit $z_r$ in the {\em tanh} model is underestimated
typically by 5--10\%, while $\Delta z$ is underestimated by much
more. We consider several different values of the assumed clumping
factor in the CDM model, and find that the $C=0$ and $C=1$ curves lie
on top of each other, while $C=10$ is only slightly different (where
the comparison is made with fixed $z_r$ and $\Delta z$ values in the
input CDM model).

\subsection{Detection limits of the global 21-cm signal}\label{datdd455}

\begin{figure*}
\begin{center}
 \hbox{
\psfig{figure=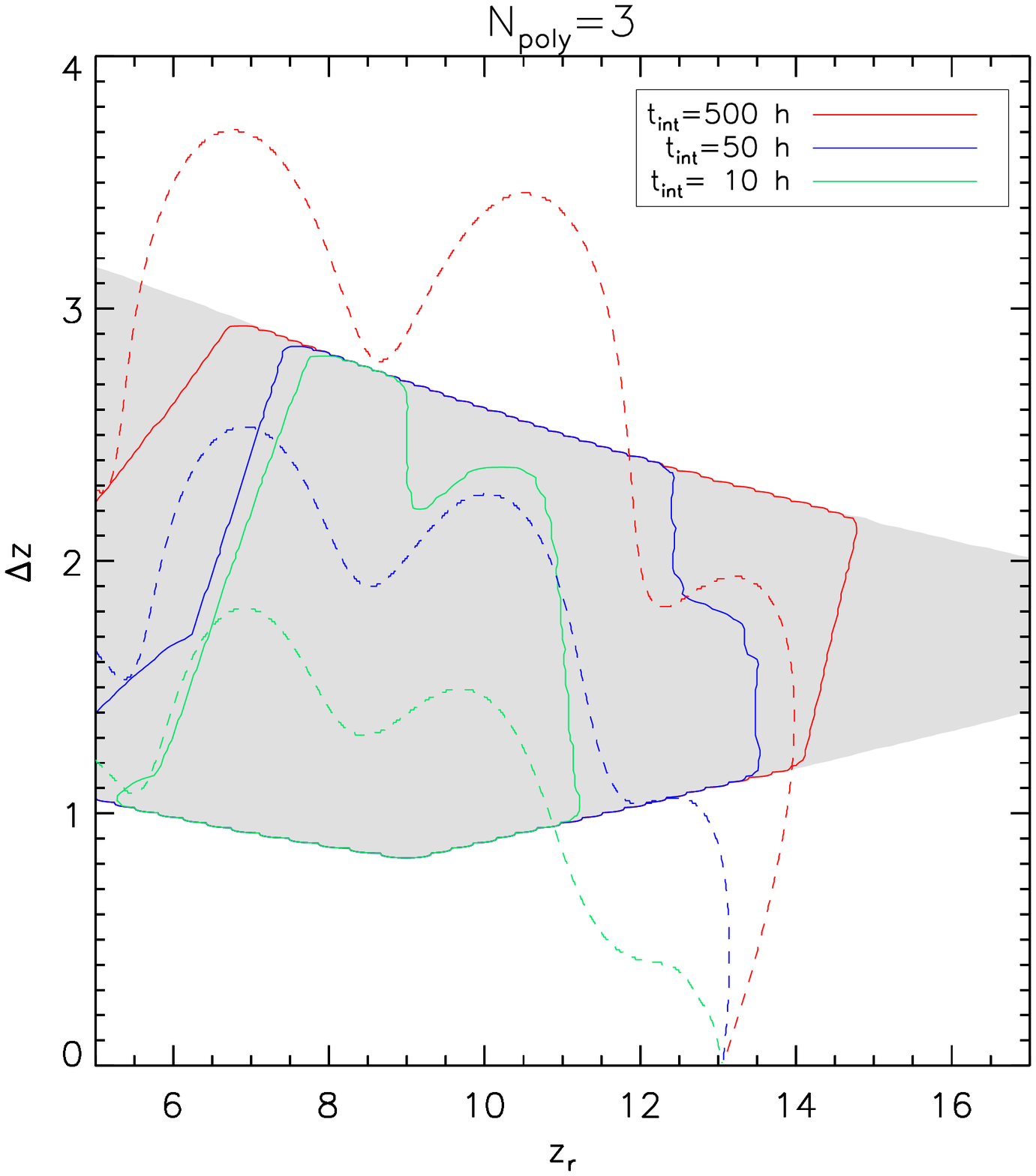,width=0.48\textwidth}
\psfig{figure=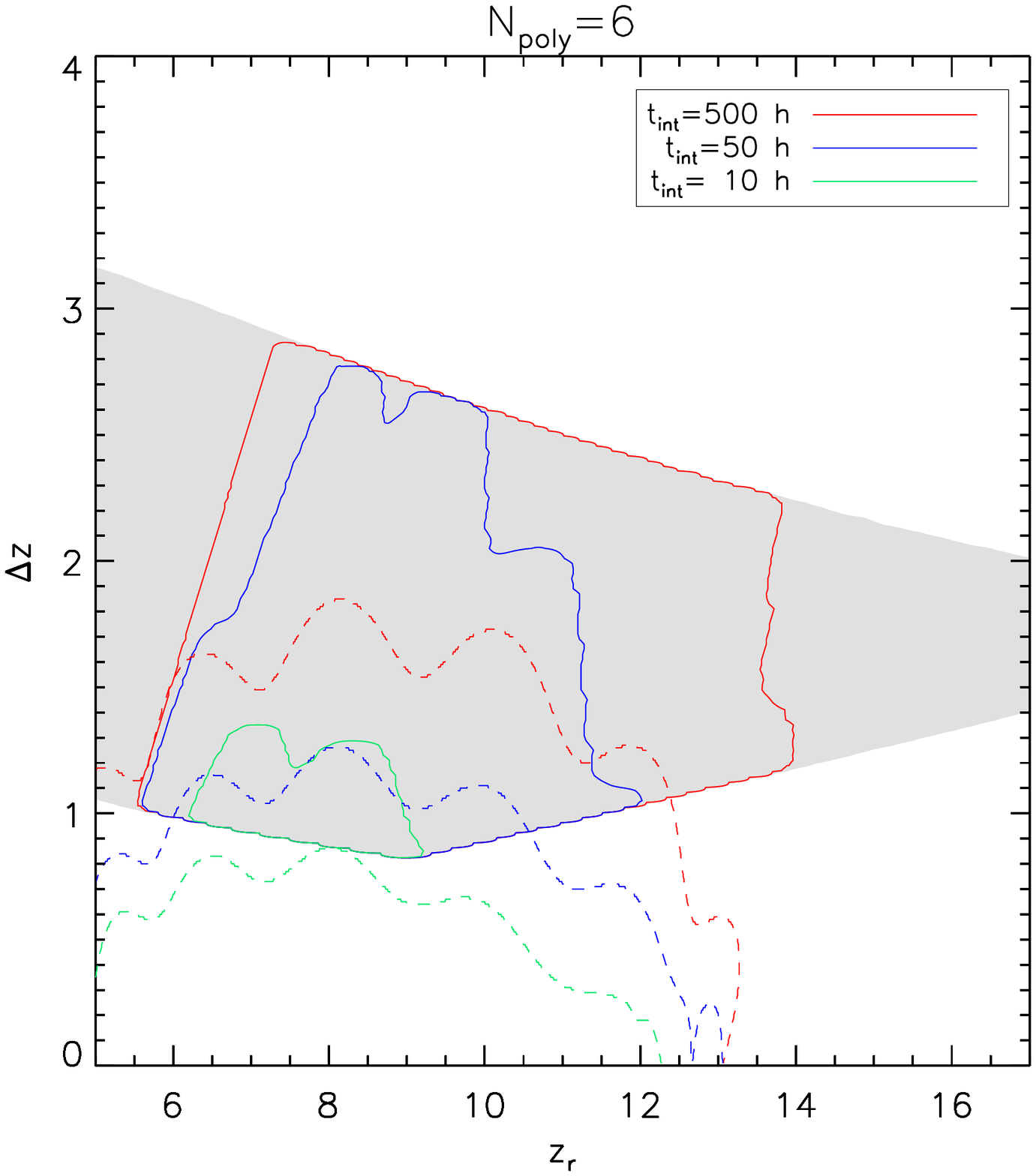,width=0.48\textwidth}
}
 \hbox{
\psfig{figure=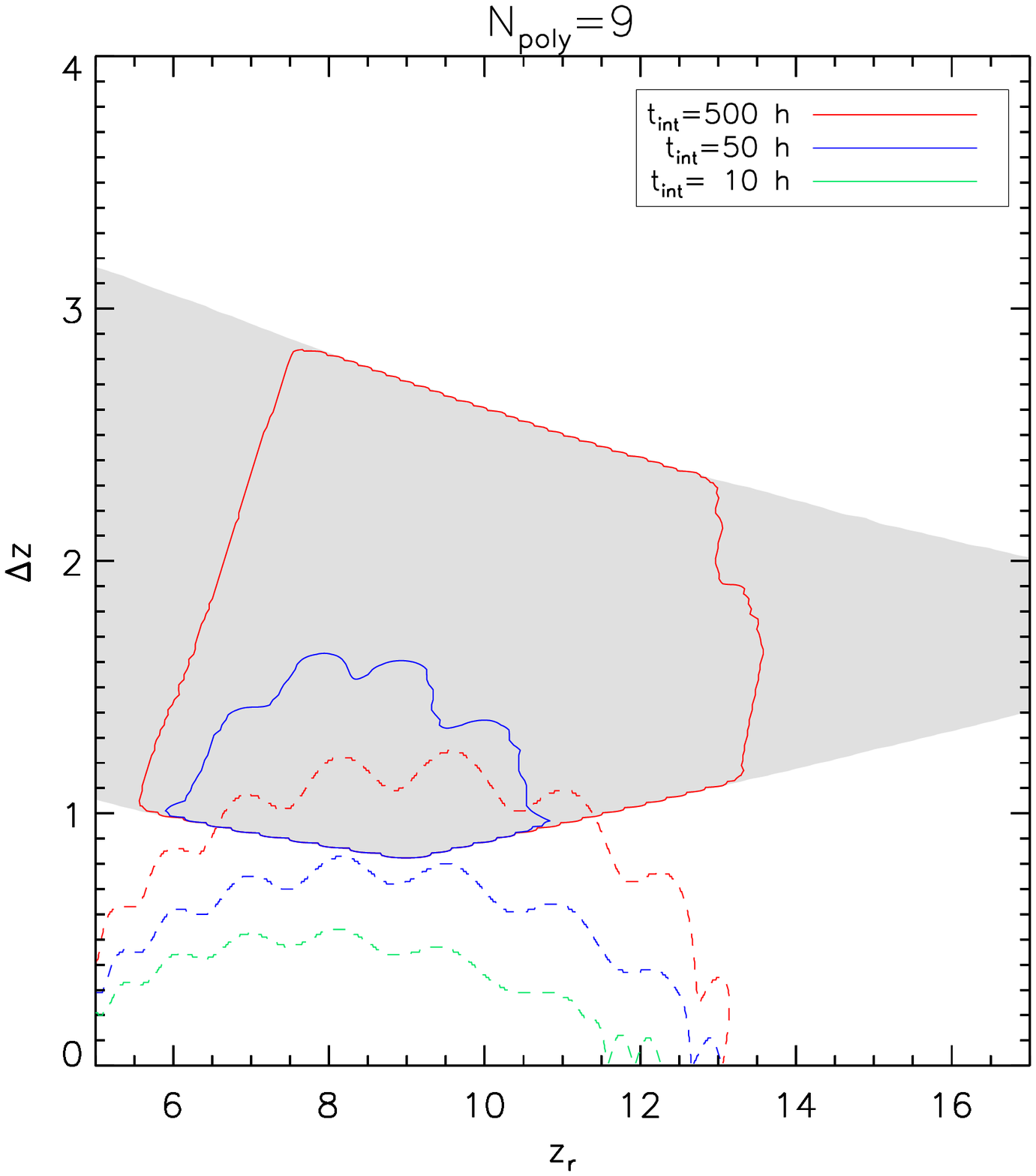,width=0.48\textwidth}
\psfig{figure=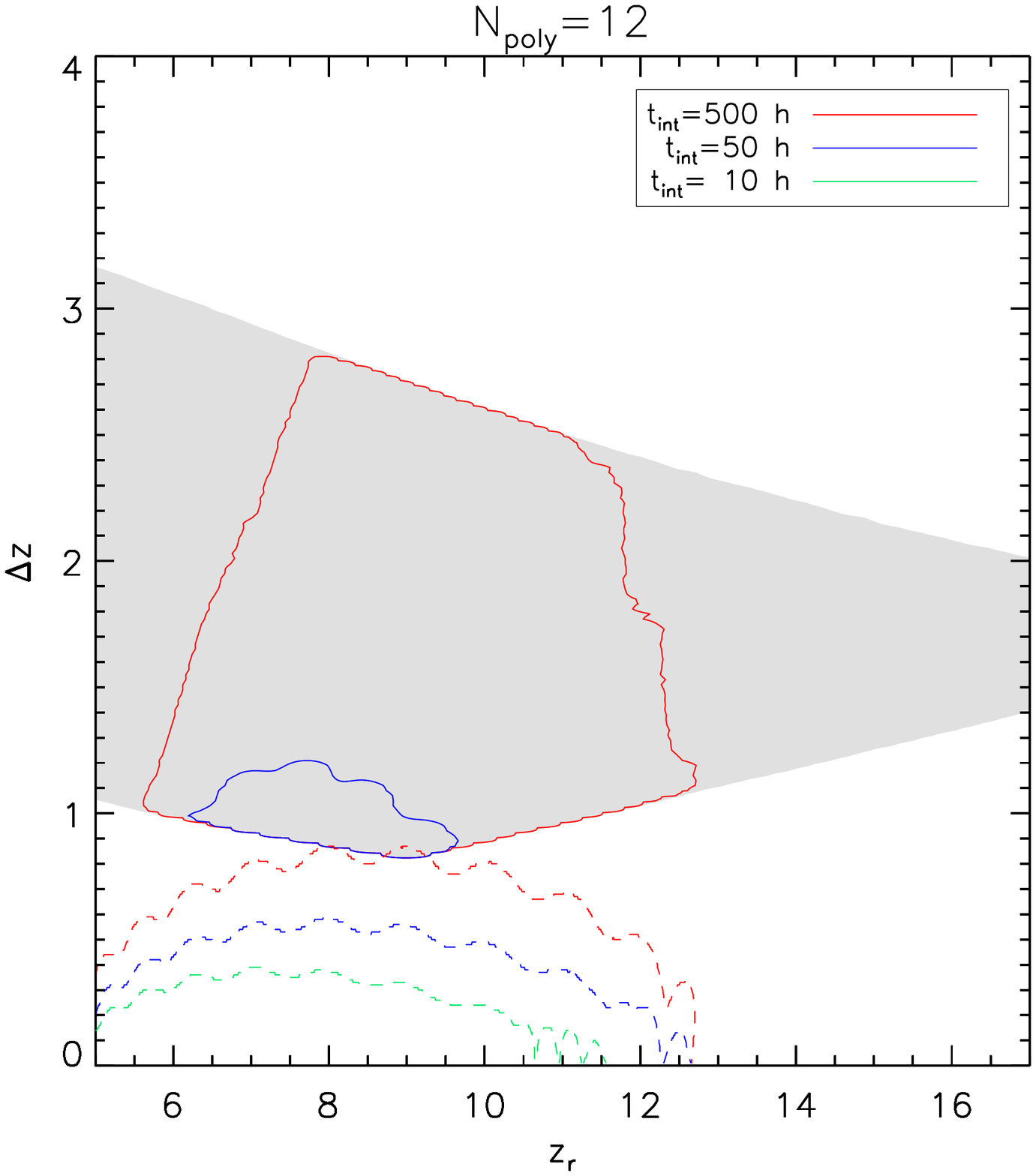,width=0.48\textwidth}
}
\caption[]{The 95\% detection region for global 21-cm experiments, 
in terms of the midpoint $z_r$ and span $\Delta z$ of reionization.
We consider polynomial order in the fit $N_{\rm poly}=\{3,6,9,12\}$ in
the various panels. We consider both the CDM model (solid curves) and
the {\em tanh} model (dashed curves), in each case for an
observational integration time $t_{\rm int}=500, 50,\, {\rm or}\, 10$ hours
(red, blue, and green, respectively, also from top to bottom). We also
show the full range of allowed values of $z_r$ and $\Delta z$ of
reionization (gray shaded area) assuming the parameter space $N_{\rm
ion}\in \{2,1000\}$, $V_{\rm c}\in \{4.5,100\}$ km/s, $C=1$.}
\label{ent246g}
\end{center}
\end{figure*}

In this section we present our main result, i.e., the experimental
sensitivity that is required to detect the global 21-cm signal, as
predicted by each of the reionization models. A range of different
results is summarized in Figure~\ref{ent246g}. First we display the
full range of allowed values of the midpoint and span of reionization
within the CDM model (gray shaded region), where the parameters
($N_{\rm ion},V_{\rm c}$) are allowed to vary over $N_{\rm ion}\in
\{2,1000\}$ and $V_{\rm c}\in \{4.5,100\}$ km/s, fixing $C=1$. 
This region reflects Figure~\ref{ent2rt2}, showing that a wide range
of $z_r$ is plausible, while the most relevant range of $z_r = 6-12$
includes some models with $\Delta z$ as low as $\sim 1$.

The figure also shows curves which delineate the 95\% detection region
for the different reionization models, for various polynomial orders
of the fit ($N_{\rm poly}=\{3,6,9,12\}$ in equation
\ref{eq:Tb2y}), and several possible values of the integration time 
$t_{\rm int}$. We define a model as detected if it is inconsistent
with a fit that does not include the reionization signal. Thus, we fit
each model signal with a foreground polynomial of a particular $N_{\rm
poly}$, and if the resulting minimum $\chi^2$ is inconsistent with zero
at greater than 95\% confidence, than that model is included within
the detection region. For the CDM model, the unrealistic sudden end to
reionization raises the $\chi^2$ values somewhat, so we reduce our
sensitivity to this by removing from the $\chi^2$ value the
contribution of the 12 MHz band centered on the end of reionization
for each model. For both models there are some oscillations in the
detection limit curves due to the degeneracy between the 21-cm signal
and the polynomial fitting of the foreground.

For the {\em tanh} model, for the same case as PL ($t_{\rm int}=500$
hours) we find significantly better prospects for detectability, with
our limits on $\Delta z$ for a given $z_r$ typically higher by a
factor of $\sim 1.5$ than their result. It is possible to constrain
some models with $\Delta z \sim 1$ even with $t_{\rm int}=10$ hours,
if $N_{\rm poly}=3$ suffices for removing the foreground and other
systematics, or with $t_{\rm int}=500$ hours if $N_{\rm poly}=9$. The
worst case of $N_{\rm poly}=12$ only allows the detection (or ruling
out) of very sharp reionization models that are probably unrealistic.

Our CDM model gives comparable constraints to the {\em tanh} model for
low values of $N_{\rm poly}$, but it is significantly more detectable
with higher-order polynomials. The rapid rise of reionization up until
its sharp end is easier to distinguish from a high-order polynomial
compared with the smooth {\em tanh} model, even after the removal (in
the CDM case) of the frequency interval right near the end of
reionization. We thus find that an interesting parameter space of CDM
models can be detected even with $N_{\rm poly}=12$, for integration
times of at least $\sim 50$ hours. 

We have not tried to indicate current constraints on reionization in
in Figure~\ref{ent246g}, to avoid overcrowding the figure, especially
since these constraints are not directly expressed in terms of $z_r$
and $\Delta z$, and the conversion to these variables would differ
somewhat between the {\em tanh} and CDM models. Roughly, for these
models, the 7-year WMAP data implies a 95\% limit of $z_r \ga 8$,
while the absorption constraints that show a high ionization fraction
at $z \sim 6.5$ imply a minimum $z_r$ that increases beyond 8 if
$\Delta z \ga 1$ (see also the Introduction and the discussion in
PL). The relation between these constraints on reionization and those
from global 21-cm measurements would change for more complex models of
reionization.

\section{Conclusions}\label{datdd5}

The aim of this paper was to investigate the possibility that global
21-cm observations during the epoch of reionization can probe the
evolution of the IGM and the physical properties of the ionizing
sources. Detecting the 21-cm signal in the presence of the large
foregrounds is challenging and it is important to explore all avenues.
While interferometric radio arrays are gearing up to measure 21-cm
fluctuations, global measurements with a single dipole experiment can
provide an independent and complementary method for detecting and/or
constraining reionization.

In order to derive quantitative predictions, we have implemented both
a previously-used toy model and a more realistic and
physically-motivated model for reionization. The first one, the {\em
tanh} model, is expressed in terms of two parameters, namely the two
main characteristics of the overall reionization process, its midpoint
$z_r$ and span $\Delta z$; the particular form of the model is merely
mathematically convenient, with no real physical significance, and it
restricts reionization to be smooth and symmetric about its midpoint.
The second model, the CDM model, is based on the standard
understanding of galaxy formation within CDM-dominated halos. It
assumes a fixed overall ionizing efficiency $N_{\rm ion}$ (number of
ionizing photons per baryon), a density clumping factor $C$ and a
minimum halo circular velocity $V_{\rm c}$ for galactic halos, and it
yields reionization models with up to 3 parameters ($C$ is relatively
minor and we typically held it fixed in the fitting). Unlike the {\em
tanh} model, the CDM model is asymmetric, with the exponentially
increasing halo abundance leading to an acceleration of reionization
in its later stages.

Despite the fact that the {\em tanh} model is a simple
parameterization that has often been used in the literature, we have
shown that it leads to substantial systematic errors if it is assumed
when fitting a 21-cm signal that is described by the CDM model. In
particular, the best-fit $z_r$ in the {\em tanh} model is
underestimated typically by 5--10\%, while $\Delta z$ is
underestimated by tens of percent. However, a sufficiently sensitive
experiment (e.g., with an integration time $t_{\rm int} > 20$ hours
for the case of a foreground polynomial of degree $N_{\rm poly}=3$)
would be able to discriminate between the CDM and {\em tanh} models
based on the $\chi^2$ value of the best-fit model.

Our main result is a detailed plot of the detection limits of global
21-cm experiments (Figure~\ref{ent246g}). We find that the CDM model
can produce quick reionization scenarios (with a redshift span $\Delta
z \sim 1$) if feedback makes large halos dominate, which then requires
a high ionizing efficiency in these halos (see also
Figure~\ref{ent2rt2}). Some of these realistically possible models can
be ruled out with 50-hour global 21-cm experiments even in the
pessimistic case where a polynomial of degree $N_{\rm poly}=12$ is
required for removing the foregrounds (or other systematic effects).
If somewhat more ambitions experiments are achievable, then a broad
range of scenarios up to $z_r \sim 12$ can be probed within our CDM
model. The smooth and symmetric {\em tanh} model is more difficult to
differentiate from the foreground polynomial, and it requires greater
integration times and lower $N_{\rm poly}$ in order to rule out for
similar reionization characteristics. 

Our conclusions are generally optimistic in terms of the possibility
for global 21-cm experiments to reconstructing the reionization
history and constrain the properties of the ionizing sources. In
particular, one-year EDGES observations may allow a remarkably precise
reconstruction. However, the polynomial degree $N_{\rm poly}$ that is
required for removing the foreground and systematic effects plays an
important role. In the most optimistic case, where $N_{\rm poly}=3$
suffices over the entire frequency range of 100--250 MHz, $1\%$ errors
on $z_r$ are achievable; they require $t_{\rm int}=51$ hours with the
{\em tanh} model or 29 hours with the CDM model (in each case with
$\Delta z=2$ as defined in that model). These, of course, are only
statistical errors, while we have shown that there can be much larger
systematic errors if the assumed reionization model cannot reproduce
the real 21-cm signal from reionization.

Indeed, our results merit some caution, since our investigation
indicates that a broader range of flexible and realistic models of
reionization should be studied before we can be confident that the
results are robust. For instance, the parameters of the model ($N_{\rm
ion}$, $V_c$, and $C$) could change with redshift due to evolving
feedbacks such as metal enrichment or the effect of photoheating on
suppressing gas accretion onto galaxies in the reionized regions (this
effect is large if reionization is initially dominated by relatively
small halos). Such an evolution could, e.g., be parameterized as in
\citet{barkana2009}; while such a complication of the model would no
doubt lead to serious partial degeneracies among the parameters,
hopefully the main characteristics of reionization would remain
measurable at high accuracy. Another possible complication that could
be added is the increasing effect of recombinations near the end of
reionization due to the optical depth of dense clumps within the
then-large ${\rm H\:II}$ bubbles \citep{FO05}. We note that
some of these effects, such as the feedback and the increasing
recombinations, should slow the progress of reionization as it nears
its end, thus rounding the shape of the CDM model and perhaps reaching
a result that is more similar to the tanh model. We plan to study
these more realistic models.

We note that in the radiometer equation we
neglected the effects of the instrumental response (or bandpass) on
both the foreground and the cosmological signal (i.e., we assumed a
flat bandpass filter). Indeed, in a real observation this would raise
three issues, but it should be possible to deal with them (Judd
Bowman, personal communication). First, our theoretical model must be
convolved with the instrumental response, in order to compare to the
observations. This can be done if the bandpass is known with sufficient
accuracy ($\sim$1\% in current experiments and expected to improve). The
foregrounds must also be convolved with the response, but as long the
instrumental response is smooth, i.e., does not introduce sharp
frequency features, the foregrounds can still be removed by low-order
polynomial fitting. Finally, another effect of a less than perfect
response is a decrease of the sensitivity of the observation with
respect to a flat receiver response. We can simply compensate for this
effect by slightly increasing the observing time (by a few tens of
percent for current experiments).

Recently, \citet{BowmanNature} reported substantial new results from
their upgraded EDGES experiment. Taking only the cleanest data out of
their observational run, they had a thermal noise level equivalent to
$t_{\rm int}=0.8$ hour (with our idealized assumptions) within their
100-200 MHz band. They also found it necessary to use $N_{\rm poly}=5$
in order to remove the foreground and systematics, and reach the
thermal noise level, within 20 MHz sub-bands in their spectrum. This
is still somewhat worse than even our conservative $N_{\rm poly}=12$
case over the full 100-250 MHz range. Even with these limitations,
\citet{BowmanNature} reached an observational milestone, namely the 
first direct observational limit on the rapidity of cosmic
reionization. In particular, using the {\em tanh} model, they set a
95\% confidence lower limit of $\Delta z > 0.06$ for the duration of
the reionization epoch. Global 21-cm experiments are still in their
infancy and clearly have a quite promising future. 

\section*{acknowledgements}
We acknowledge support by Israel Science Foundation grant 823/09. We thank Abraham Loeb and Jonathan Pritchard for useful discussions.



\newcommand{\noopsort}[1]{}

\end{document}